# Rise times of voltage pulses in NbN superconducting single-photon detectors

K.V. Smirnov[1),2),5)], A.V. Divochiy[2)], Yu.B. Vakhtomin[1),2)], M.V. Sidorova[1)], U.V. Karpova[1),2)], P.V. Morozov[2), a)], V.A. Seleznev[1),2)], A. N. Zotova[3),4)], D.Yu. Vodolazov[3),4)]

[1] *Moscow State Pedagogical University, 1 Malaya Pirogovskaya St., 119435, Moscow, Russian Federation*

[2] *CJSC "Superconducting nanotechnology" (Scontel), 5/22-1 Rossolimo St., 119021, Moscow, Russian Federation*

[3] *Institute for Physics of Microstructure, RAS, 603950, Nizhny Novgorod, GSP-105, Russian Federation*

[4] *Lobachevsky State University of Nizhny Novgorod, 23 Gagarin Avenue, 603950 Nizhny Novgorod, Russian Federation*

[5] *National Research University Higher School of Economics, Moscow Institute of Electronics and Mathematics, 34 Tallinskaya St., 109028, Moscow, Russian Federation*

We have found experimentally that the rise times of voltage pulses in NbN superconducting single photon detectors increase nonlinearly with increasing detector length. We fabricated superconducting single photon detectors based on NbN thin films with a meander-like sensitive region of area from $2 \times 2$ μm$^2$ to $11 \times 11$ μm$^2$. The effect is connected with the dependence of the detector resistance, which appears after photon absorption, on its kinetic inductance and hence on detector length. This conclusion is confirmed by our calculations in the framework of the two-temperature model.

---

[a)]Author to whom correspondence should be addressed. Electronic mail: morozov@scontel.ru

Since the discovery of single photon detection in the visible and near-infrared band using superconducting nanostructures [1], a new class of single photon detectors was created, namely the superconducting-nanowire single-photon detector (SSPD or SNSPD), which have been successfully used in many applications [2,3]. Nevertheless, there are currently a number of ongoing investigations improving these superconducting detectors and studying their mechanisms. The appearance of the resistive region in superconductors under photon absorption, the further relaxation of hot carriers, and the recovery of superconductivity is a widely studied process because it determines many parameters of the SSPD such as dead time, limiting counting rate, and its temporal resolution. The current work presents the results of experimental and theoretical investigations of voltage-pulse rise time arising in a superconductor strip under the absorption of single IR photons.

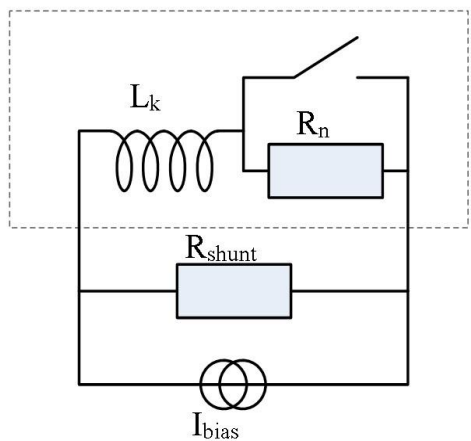

FIG. 1. Equivalent circuit of the SSPD according to Kerman et al., Appl. Phys. Lett. **88**, 111116 (2006).

In previous work [4], we investigated the voltage pulse and represented the SSPD as a combination of several electrical elements, two of which have kinetic inductance $L_k$ and resistance $R_n$ that occur in a superconducting strip when a photon is absorbed (Fig. 1). We derived analytical expressions for the rise time ($\tau_{rise}$) and attenuation ($\tau_{rise}(L_k)$) of the voltage pulse at the detector, $\tau_{rise}=L_k/(R_{shunt}+R_n)$ and $\tau_{fall}=L_k/R_{shunt}$, respectively, by solving the Kirchhoff equation for the electrical circuit of the detector, assuming $R_n$ is constant. These estimations agreed well with experimental measurements of the decay time of the voltage pulse for the SSPD. However, we did not investigate the rise time of the voltage pulse. We obtained experimental results for $\tau_{rise}(L_k)$, which were not consistent with the proposed model [4]. Usually SSPDs have an active meander-like region of area 100–200 μm$^2$. Therefore, reducing the size of the SSPD active region to 5–10 μm$^2$, we expect to see a significant reduction both in the rise and fall times of the output pulse. Indeed, the reduction fall time is completely described by the decrease in kinetic inductance in accordance with τ$_{fall}$ above. However, our investigations show that the rise time of the voltage pulse for the SSPD decreases with decreasing kinetic inductance more slowly than that predicted by τ$_{rise}$ above and is a nonlinear function of L$_k$. Therefore, we decided to investigate the behavior of the pulse rise time for the SSPD in more detail.

To measure τ$_{rise}$(L$_k$), we fabricated superconducting single photon detectors based on NbN thin films with a sensitive region following a meander-like pattern [5] with fill factor of ~0.55. The superconducting NbN strip covers an area from 2.2 μm$^2$ to 11.11 μm$^2$. The characteristics of the fabricated detectors are listed in Table I. We used two types of substrate on which to deposit the NbN films. The first type is Si with an additional Si$_3$N$_4$ layer, which was deposited by the CVD technique and has a thickness of 150 nm. The second type is Si with a SiO$_2$ layer, which was created by thermal oxidation. The SiO$_2$ layer has a thickness of 200 nm. All batch-fabricated detectors with a variety of long superconducting strips have the same value of superconducting transition temperature T$_c$=8.5 K and very close values of critical currents I$_c$ of ~8 μA at a temperature of 4.2 K. The calculated resistance of each detector is fully consistent with the topology and surface resistance of its superconducting film, which was measured after the deposition of the NbN film.

TABLE I. Characteristics of NbN films.

| Type | Substrate | NbN thickness | NbN width | Surface resistance | Size of active area | NbN length | Kinetic inductance |
|---|---|---|---|---|---|---|---|
| 1 | Si+Si$_3$N$_4$ | 5 nm | 110 nm | 430–470 Ohm/□ | 2.2 μm$^2$ | 20 μm | 20 nH |
| 2 | CDV, | | | | 3.3 μm$^2$ | 46 μm | 46 nH |

| 3 | 150 nm | | | | 5·5 μm² | 128 μm | 128 nH |
|---|---|---|---|---|---|---|---|
| 4 | | | | | 7·7 μm² | 250 μm | 250 nH |
| 5 | | | | | 11·11 μm² | 605 μm | 605 nH |
| 6 | Si+SiO$_2$ thermal oxidation, 200 nm | | | | 2·2 μm² | 20 μm | 20 nH |
| 7 | | | | | 3·3 μm² | 46 μm | 46 nH |
| 8 | | | | | 5·5 μm² | 128 μm | 128 nH |
| 9 | | | | | 7·7 μm² | 250 μm | 250 nH |

The NbN detectors were cooled by liquid helium to a temperature of 4.2 K. The radiation from the laser source (FHS1DO2, λ=1.55 μm) was delivered to the sample via a standard single-mode fiber (SMF 28e, Corning Inc., Corning, NY). The voltage pulse ΔU that arises under photon absorption was amplified by a cascade of two wideband amplifiers (RFCOMP HD27067, bandwidth: 20–4000 MHz, gain 25 dB, HD Communications, Ronkonkoma, NY) and analyzed using a high-frequency oscilloscope (DPO70404c, Tektronix, Beaverton, OR). We tested the full high-frequency chain including the CPW line, coaxial cable, bias-T adapter, and two amplifiers for short-pulses distortion. The test shows that the amplitude frequency characteristic of the chain allows pulsed signals to pass with a rise/fall time as low as 50 ps without significant distortion. All measurements were performed at bias currents $I_b$ which correspond to the level of a dark counts ~10 cps ($I_b$~0.8–0.9 $I_c$). Also the power of the laser source corresponds to a count of ~$10^6$ Hz. The rise time $\tau_{rise}(L_k)$ was determined from the dependence ΔU(t) in Fig. 2 on the time interval between levels one-tenth and nine-tenths of $U_{max}$.

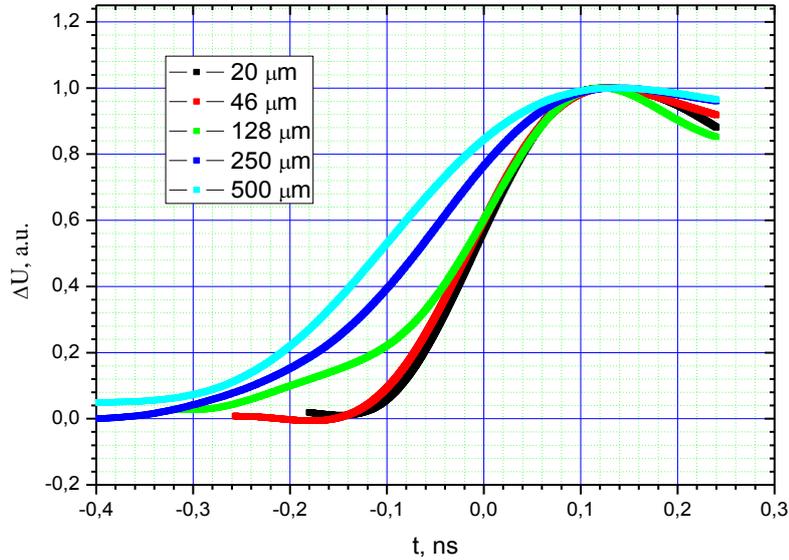

FIG. 2. Time dependence of $\Delta U(t)$ for detectors of different strip lengths. The maximum value $\Delta U$ is normalized to unity.

The experimental results for $\tau_{rise}$ are presented in Fig. 3 (solid symbols). One can see that the $\tau_{rise}(L_k)$ are nonlinear for all studied detectors. Physically, such a nonlinear dependence $\tau_{rise}(L_k)$ can be understood in the following way. The absorbed photon locally heats the superconductor and, if locally the current exceeds a temperature-dependent critical current $I_c(T)$, the Joule dissipation starts to heat the superconductor and the normal region (domain) spreads along the strip. The resistance $R_n$ of the detector increases and the current deviates from the superconductor and flows via the shunt resistor (see Fig. 1). With larger kinetic inductance $L_k$, the time needed for current to avoid the superconductor is longer, and the normal domain that it produces is larger in size with larger $R_n$. When the current through the detector becomes smaller than some critical value (at this level of current, the heat flowing to the substrate exceeds Joule heating), the normal domain starts to shrink and when the voltage drop via the detector falls below that via the shunt, the current returns to the detector. At this instant, the voltage on the shunt (which is measured in the experiment and is given in Fig. 2) reaches its maximal value $U_{max}$. If we use a simple estimation of $\tau_{rise} \sim L_k/R_n(L_k)$ then it is easy to see that because of the dependence of $R_n$ on the length, and hence $L_k$, the dependence $\tau_{rise}(L_k)$ should be nonlinear.

To support this rough argument, we calculated the thermoelectric response of the SSPD after photon absorption. In our model, we assume that the absorbed photon heats locally the electrons and phonons. We also assume that the deviation of electrons and phonons from equilibrium could be described by a two-temperature (2T) model. In this model, electrons and phonons have their own temperatures ($T_e$ and $T_{ph}$) different from the temperature of the substrate $T_{sub}$ and their temporal and spatial evolution can be found from the solution of two coupled equations,

$$C_e \partial T_e/\partial t = \partial/\partial x(\kappa_n \partial T_e/\partial x) + \rho(x,T_e)(I_d/wd)^2 - 96\zeta(5)N(0) k_B^2(T_e^5 - T_{ph}^5)/(\tau_0 T_c^3), \qquad (1)$$

$$\partial T_{ph}^4/\partial t = -(T_{ph}^4 - T_{sub}^4)/\tau_{esc} + 360\zeta(5)\gamma/\pi^4 \bullet (T_e^5 - T_{ph}^5)/(\tau_0 T_c). \qquad (2)$$

This pair of equations can be derived from the kinetic equations for electrons and phonons in the pure metal if one seeks the solution for electron and phonon distribution functions in the form $f_e(E)=1/(\exp(E/k_B T_e)+1)$ and $f_{ph}(E)=1/(\exp(E/k_B T_{ph})-1)$. For small deviations from equilibrium $T_e - T_{sub} \ll T_{sub}$ and $T_{ph} - T_{sub} \ll T_{sub}$ and spatial uniformity, Eqs. (1) and (2) reduce to the equations presented in Ref. [6]. In Eq. (2), we neglect the diffusion of phonons because the group velocity of phonons (speed of sound $v_s$) is much smaller in comparison with that of electrons (Fermi velocity $v_F$). From Eq. (1), the familiar temperature dependence of the electron–phonon relaxation rate $1/\tau_{e\text{-}ph} \sim T_e^3$ is obeyed for electrons with energy less than $k_B T_e$.

The effective temperature approach should be correct for dirty NbN superconductors having a short electron–electron inelastic relaxation rate $\tau_{e\text{-}e} \sim 7$ ps [7], which is much smaller than $\tau_{rise}$ for any length of those detectors studied. In Eqs. (1) and (2) $\kappa_n(T_e \geq T_c) = 2D\pi^2 N(0)k_B^2 T_e/3$ is the electron heat conductivity in the normal state (D is a diffusion coefficient), $\tau_{esc}$ is the escape time of the non-equilibrium phonons into the substrate, $T_c$ the critical temperature of the superconductor, $\tau_0$ the characteristic time introduced in [8] (it couples to the electron–phonon collision integral in the electron kinetic equation), parameter $\gamma=1.6\pi^2 C_e(T_c)/C_{ph}(T_c)$ (the ratio $\gamma/\tau_0$ couples to the phonon–electron collision integral in the phonon kinetic equation) is proportional to the ratio of the heat capacities per unit volume for electrons $C_e(T \geq T_c)=2\pi^2 k_B^2 N(0) T_e/3$ and phonons $C_{ph}=12\pi^2(T_{ph}/T_D)^3 k_B n_{ion}/5$ at $T_e==T_{ph}=T_c$, $N(0)$ the density of states of electrons per spin at the Fermi level, $T_D$ the Debye temperature, $n_{ion}$ the ion concentration, and $\zeta(5) \sim 1.04$.

The term proportional to $I_d^2$ in Eq. (1) describes the Joule heating in the detector where the transport current exceeds the temperature-dependent critical current $I_c(T_e)$ ($w$ is the width of the strip, $d$ its thickness). The temperature and coordinate-dependent resistivity $\rho(x,T_e)$ is equal to the normal state resistivity $\rho_n$ in the area where $T_e(x)>T_c$, $\rho(x,T_e)=\rho_n(1-I_c(T_e)/I_d)$ at $T_e(x)<T_c$ and $I_d>I_c(T_e)$, and $\rho(x,T_e)=0$ when $I_d<I_c(T_e)$.

In dirty metals, the electron–phonon interaction is modified through the presence of impurities, which provides the modification of the electron–phonon relaxation rate. We consider the case when $1/\tau_{e\text{-}ph} \sim T_e^2$ [9], which is close to the temperature dependence $1/\tau_{e\text{-}ph} \sim T_e^{1.6}$ experimentally observed in NbN [10]. Theoretically, a $1/\tau_{e\text{-}ph} \sim T_e^2$-dependence stems from the energy-dependent electron–phonon coupling constant (which is inversely proportional to the energy in this case, which leads to a modification to the Eliashberg function [11]) in the electron–phonon collision integral. It produces a modification of the last two terms on the right-hand side of Eqs. (1) and (2), which describe the coupling between electrons and phonons,

$$C_e \partial T_e/\partial t = \partial/\partial x(\kappa_n \partial T_e/\partial x) + \rho(x,T_e)(I_d/wd)^2 - 4\pi^4 N(0) k_B^2 (T_e^4 - T_{ph}^4)/(15\tau_0 T_c^2), \qquad (3)$$

$$\partial T_{ph}^4/\partial t=-(T_{ph}^4-T_{sub}^4)/\tau_{esc}+\gamma(T_e^4-T_{ph}^4)/\tau_0. \qquad (4)$$

Equations (1)–(4) are strictly valid only in the normal state. To describe the temporal and spatial evolution in regions of the detector where $T_e<T_c$, we use all four equations but with $C_e(T<T_c)=3.86\bullet 2\pi^2 k_B^2 N(0)T_e(1-0.37\bullet T_c/T_e)/3$ (which gives a reasonable approximation of the exact $C_s(T_e)$ at $0.4<T_e/T_c<1$) and $\kappa_s(T<T_c)=k_n T/T_c$ and is a widely used approximation for moderate temperatures ($0.3<T_e/T_c<1$) (see for example Ref. [12]). In the superconducting state the last terms on the right-hand side of Eqs. (1)–(4) also should be modified because of the presence of the energy gap in the spectrum of electrons, which leads to their reduction. In our case with strong disequilibrium, $T_e$ and $T_{ph}$ exceed $T_c$ in the largest region of the resistive/normal domain (at least in detectors with large length L) and we do not expect large effects from such a modification.

Together with Eqs. (1)–(4), we also solve the equation corresponding to the equivalent scheme for the superconducting detector (Fig. 1),

$$L_k \frac{\partial I_d}{\partial t} = (I-I_d)R_{shunt} - I_d R_n, \qquad (5)$$

where $L_k=\mu_0\lambda_L^2 L/(wd)$ is the kinetic inductance of the superconductor (here $\lambda_L$ is the London penetration depth, L the length, *w* the width and *d* the thickness of the superconducting strip, $\mu_0$ is a magnetic constant), $R_{shunt}=50$ Ohm is the shunt resistance, $R_n$ the resistance of the detector (in our model, it is proportional to the length of the superconductor where $I_d>I_c(T_e)$).

We consider a one-dimensional model because even for the shortest detectors, rise time exceeds 140 ps (see Fig. 3), which is much larger than the diffusion time of non-equilibrium electrons across half of the strip $\tau_{diff}\sim(w/2)^2/4D\sim15$ ps with width=110 nm and D=0.5 cm$^2$/s. In calculations, we started from the initial state with $T_e=T_{ph}=T_{spot}>T_{sub}$ in the region with length equal to the superconducting coherence length ξ (referred to as the initial resistive domain) and $T_e=T_{ph}=T_{sub}$ in the rest of superconductor. We model the temperature dependence of the critical current $I_c(T_e)$ using the Bardeen approximation $I_c(T_e)=I_c(0)(1-(T_e/T_c)^2)^{3/2}$, where $I_c(0)$ was found from the experimental $I_c(4.2K)=8.2$ μA. Parameter γ=9 is estimated using $n_{ion}=4.6*10^{22}$ cm$^{-3}$ calculated from the density of NbN material ρ=8.2 g/cm$^3$. The rest of the parameters have the following values typical for our NbN detectors: $T_c$=8.5 K, D=0.5 cm$^2$/s, $R_{sq}=\rho_n d$=450 Ohm, $\lambda_L(0)$=540 nm, w=110 nm, d=5 nm, N(0)=25.5 nm$^{-3}$eV$^{-1}$. In simulations, as in the experiment, we identify the rise time as the time interval needed for the voltage through the shunt to grow from 0.1•$U_{max}$ to 0.9•$U_{max}$.

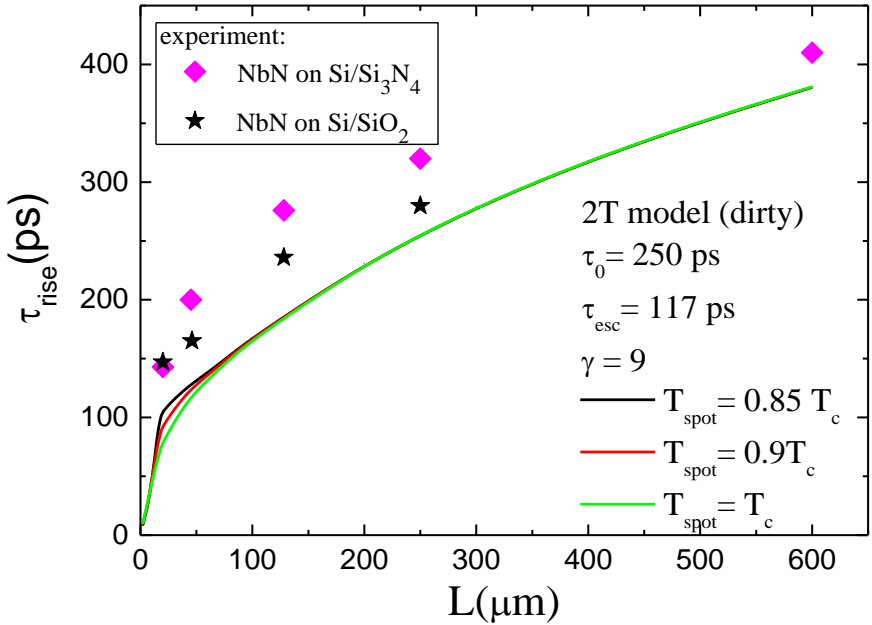

FIG. 3. Experimental and theoretical (calculated in the 2T model for a dirty metal) rise time of the voltage pulse in NbN SSPD as a function of strip length. The different colored curves correspond to different temperatures inside the initial resistive domain. For chosen $\tau_0$, $\tau_{esc}$, and γ the theoretical retrapping current coincides with experimental data both at T=4.2 K and 6.9 K with accuracy better than 3% (see Fig. 4).

Equations (1)–(4) contain two unknown parameters (if we assume that our estimation of $\gamma=9$ is correct): $\tau_0$ and $\tau_{esc}$. The time $\tau_0$ can be expressed via the energy relaxation time $\tau_E(T)$ [13] for pure metals at $T=T_c$: $\tau_0=720\zeta(5)\tau_E(T_c)/\pi^2\sim76\tau_E(T_c)$ (or via the electron–phonon relaxation time $\tau_0=14\zeta(3)\tau_{e-ph}(T_c)\sim16.8\tau_{e-ph}(T_c)$) and for dirty metals $\tau_0=8\pi^2\tau_E(T_c)/5\sim15.8\tau_E(T_c)$ ($\tau_0\sim9.8\tau_{e-ph}(T_c)$). Measurements of $\tau_E(T_c)$ in NbN [10] gives $\tau_E(T=8.5\ K)\sim16$ ps, which provides an expected value of $\tau_0\approx1200$ ps for pure metals and $\tau_0\approx250$ ps for dirty ones.

From another perspective, we also have the current–voltage characteristics of the detector with L=605 µm (see Fig. 4) measured in the voltage stabilization mode. From these curves, we extracted the retrapping current $I_r$ [12] and compared it with results of the 2T model. We find that in the 2T model for dirty metals $I_r$ depends on the sum $(\tau_0+\gamma\tau_{esc})$. This result can be understood if we consider a stationary solution of Eqs. (3) and (4) at relatively large voltages, when the normal domain is large and $\partial T_e/\partial x\sim0$ in the center of the domain. After some simple algebra, we find $T_{e,center}$

$$T_{e,center}^4=T_{sub}^4+\rho_n(I/wd)^2(\tau_0+\gamma\tau_{esc})T_c^2 15/(4\pi^4 N(0)k_B^2). \tag{6}$$

With the help of Eq. (6), the retrapping current is estimated as the current at which $T_{e,center}=T_c$ (from numerical solutions, we find that $T_{e,center}$ varies slightly with voltage but it is indeed close to $T_c$ at large voltages where $I\approx I_r$)

$$I_r^2\approx 4\pi^4 N(0)(k_B T_c)^2(wd)^2(1-T_{sub}^4/T_c^4)/(15\rho_n(\tau_0+\gamma\tau_{esc})T_c^2). \tag{7}$$

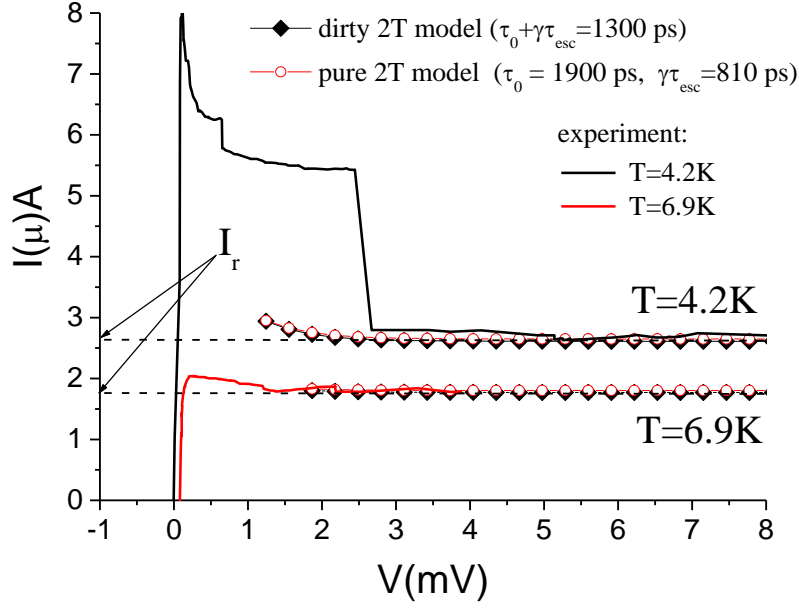

FIG. 4. Current–voltage characteristics of the meander with L=605 μm measured in the voltage stabilization mode (solid curves) and found from the numerical solution (symbols). Arrows indicate the position of the retrapping current $I_r$ at temperatures 4.2 K and 6.9 K.

Good fitting of the experimental $I_r$ occurs at $\tau_0 +\gamma\tau_{esc}\approx1300$ ps (see Fig. 4). For the 2T model for pure metals $I_r$ depends on $\tau_0$ and $\gamma\tau_{esc}$ in a more complicated manner, but the experimental $I_r$ also can be fitted as one can see from Fig. 4.

Taking into account our theoretical estimations for $\gamma=9$ and $\tau_0=250$ ps (for dirty metals) or $\tau_0=1200$ ps (for pure ones), we find $\tau_{esc}\approx115$–117 ps depending on the model. This value for $\tau_{esc}$ is slightly larger than that found in the literature $\tau_{esc}\approx80$–90 ps for NbN films with similar thickness and substrate [14,15]. We ascribe this difference to our estimation of $\gamma$, for which we have used oversimplified models for the calculation of heat capacities of electrons and phonons.

With these parameters we calculated rising time of the voltage pulse as a function of the length of the detector. In Fig. 3, we present results of our calculations using the 2T model for dirty metals, together with the experimental data (the results for pure metal practically coincide with those for dirty metal and hence are not shown). One can see that at L ~> 20 μm the theoretical $\tau_{rise}$ increases nonlinearly with increasing L, which is a consequence of the length-dependent size of the resistive domain and $R_n$. In detectors with L<~20 μm, the current through the detector decreases too fast (when the photon-induced resistive domain appears there) and the Joule dissipation cannot heat the detector strongly and size of the resistive domain is determined mainly by the initial size of the domain. In our model, it does not depend on the length of the detector and $\tau_{rise}$ increases nearly linearly with L at small L (see Fig. 3). Because in our model the resistance of the initial domain depends on its initial temperature $T_{spot}$ (via temperature dependent resistivity $\rho(T_e,I_d)$), it leads to a dependence of $\tau_{rise}$ on $T_{spot}$, but only for small lengths; at large lengths, the size of the resistive domain is determined mainly by Joule dissipation and $\tau_{rise}$ weakly depends on $T_{spot}$. Note that for the chosen parameters and chosen model for $\rho(T_e,I_d)$, we did not find in theory 'latching' [16] of the detector at any L. Experimentally, 'latching' was also not observed even for the detector with L=20 μm.

Better *quantitative* agreement between theory and experiment at large lengths (L>100 μm) is obtained if one uses smaller $\tau_0$ and larger $\tau_{esc}$, but it does not lead to better agreement at smaller lengths. Moreover, because at small L the rise time depends on the temperature of the initial domain (and/or its length), it puts into doubt the use of the present model for quantitative comparison with experiment. One needs to consider the evolution of the temperature in the two-dimensional strip just after photon absorption and use a more appropriate model for Joule dissipation in the regions of the superconductor where $T_e<T_c$ to study correctly the initial stage of the resistive domain growth, which gives the large contribution to $\tau_{rise}$ in small length detectors.

To conclude, our experimental measurements and theoretical calculations in the two-temperature model framework demonstrate a nonlinear increase in rise time of the voltage pulse in NbN SSPD with increasing length. We argued that the effect found is connected with an increasing length of the resistive domain in the detector when L increases.

D.Yu.V. acknowledges the support from the Russian Foundation for Basic Research. K.V.S. acknowledges financial support from the Ministry of Education and Science of the Russian Federation (Contract No. 3.2655.2014/K).